\documentclass[aps,twocolumn,showpacs,preprintnumbers,nofootinbib,prd,superscriptaddress,groupedaddress,10pt]{revtex4-1}
\usepackage[utf8]{inputenc}
\usepackage{graphicx,amssymb,amsmath,amsthm,amsfonts,epsfig,times,natbib}

\usepackage[usenames,dvipsnames]{color}
\usepackage{epstopdf}
\usepackage{amsmath,amssymb}
\usepackage{aas_macros}
\usepackage{tensor}
\usepackage{mathtools}
\usepackage{amsbsy}
\usepackage{bm,url}
\usepackage[linktocpage]{hyperref}

\usepackage[scaled]{beramono}
\usepackage[T1]{fontenc}

\definecolor{oxfordblue}{rgb}{0.0, 0.13, 0.28}
\definecolor{burgundy}{rgb}{0.5, 0.0, 0.13}
\definecolor{darkolivegreen}{rgb}{0.33, 0.42, 0.18}
\definecolor{darkblue}{rgb}{0,0,0.5}
\definecolor{richcarmine}{rgb}{0.84, 0.0, 0.25}
\definecolor{darkblue}{rgb}{0,0,0.5}
\definecolor{venetianred}{rgb}{0.78, 0.03, 0.08}
\definecolor{skobeloff}{rgb}{0.0, 0.48, 0.45}
\hypersetup{colorlinks=true, citecolor=darkblue, linkcolor=darkblue,
urlcolor = darkblue}

\newcommand{\ben}{\begin{enumerate}}
\newcommand{\een}{\end{enumerate}}

\def\be{\begin{equation}}
\def\ee{\end{equation}}
\def\bea{\begin{eqnarray}}
\def\eea{\end{eqnarray}}

\newcommand{\beq}{\begin{eqnarray}}
\newcommand{\eeq}{\end{eqnarray}} 
\newcommand{\ba}{\begin{align}}
\newcommand{\ea}{\end{align}}

\begin{document}

\title{The stochastic gravitational-wave background in the absence of horizons}

\author{
Enrico Barausse$^{1}$,
Richard Brito$^{2}$,
Vitor Cardoso$^{3,4}$,
Irina Dvorkin$^{2}$,
Paolo Pani$^{5}$
}

\affiliation{${^1}$ Institut d'Astrophysique de Paris, CNRS \& Sorbonne
 Universit\'es, UMR 7095, 98 bis bd Arago, 75014 Paris, France}
\affiliation{${^2}$ Max Planck Institute for Gravitational Physics (Albert Einstein Institute), Am M\"{u}hlenberg 1, Potsdam-Golm, 14476, Germany}
\affiliation{${^3}$ CENTRA, Departamento de F\'isica, Instituto Superior
T\'ecnico, Universidade de Lisboa, Avenida Rovisco Pais 1,
1049 Lisboa, Portugal}
\affiliation{${^4}$ Perimeter Institute for Theoretical Physics, 31 Caroline Street North
Waterloo, Ontario N2L 2Y5, Canada}
\affiliation{${^5}$ Dipartimento di Fisica, ``Sapienza'' Universit\`a di Roma \& Sezione INFN Roma1, Piazzale Aldo Moro 5, 00185, Roma, Italy}

\begin{abstract}
Gravitational-wave astronomy has the potential to explore one of the deepest
and most puzzling aspects of Einstein's theory: the existence of black holes. 
A plethora of ultracompact, horizonless objects have been proposed to arise in models inspired by quantum gravity.
These objects may solve Hawking's information-loss paradox and the singularity problem associated with black holes,
while mimicking almost all of their classical properties. They are, however, generically unstable on relatively short timescales.
Here, we show that this ``ergoregion instability'' leads to a strong stochastic background of gravitational waves,
at a level detectable by current and future gravitational-wave detectors. The absence of such background in the first observation run of Advanced LIGO already imposes the most stringent limits to date on black-hole alternatives, showing that certain models of ``quantum-dressed'' stellar black holes can be at most a small percentage of the total population. The future LISA mission will allow for similar constraints on supermassive black-hole mimickers.
\end{abstract}

\maketitle
\section{Introduction}
According to General Relativity and the Standard Model of particle physics, 
dark and compact objects more massive than $\approx3M_{\odot}$ must be black holes (BHs)~\cite{PhysRevLett.32.324}. These are characterized by
an event horizon, causally disconnecting the BH interior from its exterior, where observations take place.
In classical gravity, BHs can form from gravitational collapse~\cite{PhysRevLett.14.57}, providing a sound and compelling theoretical support for their existence.
When quantum effects are included at semiclassical level, however, BHs are not completely ``dark'', but evaporate 
by emitting thermal black-body radiation~\cite{Hawking:1974sw}. For astrophysical BHs, evaporation is negligible.
Therefore, BHs are commonly accepted to exist -- with masses in the ranges from $\sim 10 M_\odot$ to $\sim 60 M_\odot$
(and perhaps larger mass), and from $\sim 10^{5} M_\odot$ to $\sim 10^{10} M_\odot$  -- and 
play a fundamental role in astronomy and astrophysics~\cite{Narayan:2005ie}.

It is sometimes not fully appreciated that BHs are truly ``holes'' in spacetime, where time ``ends'' and inside which the known laws of classical physics break down~\cite{Hawking:1976ra}. 
Furthermore, the classical concept of an event horizon seems at clash with quantum mechanics, and the very existence of BHs leads to unsolved conundra such as information loss~\cite{Hawking:1976ra}. Thus, in reality, the existence of BHs is an outstanding
event, for which one should provide equally impressive evidence~\cite{Cardoso:2016rao,Cardoso:2017cfl,Cardoso:2017cqb}. 
Over the last decades several alternatives and arguments have been put forward according to which --~in a quantum theory~-- 
BHs would either not form at all, or would just be an ensemble of horizonless quantum states~\cite{Mazur:2004fk,Mathur:2005zp,Mathur:2008nj,Barcelo:2015noa,Danielsson:2017riq,Berthiere:2017tms,Cardoso:2017njb}.

From a theoretical standpoint, BHs therefore lay at the interface between classical gravity, quantum theory and thermodynamics, and understanding their nature may provide a portal to quantum gravity or other surprises.
In the formal mathematical sense, it is impossible to ever show that BHs exist, 
since in General Relativity their definition requires knowledge of the whole spacetime, including the future~\cite{PhysRevLett.14.57}. However, the newborn gravitational-wave (GW) astronomy allows us to constrain alternatives to BHs to unprecedented level. GW detectors can rule out a wide range of models, through observations of inspiralling binaries or the relaxation of the final object forming from a merger~\cite{Gair:2012nm,Cardoso:2016ryw,Cardoso:2016rao,Maselli:2017cmm,Cardoso:2017cfl,Cardoso:2017cqb}.
Here, we explore one significant effect that follows from the absence of the most salient feature of a BH, the event horizon. 
We will show that compact, horizonless spinning geometries would fill the universe with a background of GWs detectable by current and future instruments, through a classical process known as the ``ergoregion instability''~\cite{1978CMaPh..63..243F,Brito:2015oca}.
\section{Ergoregion instability}
In Einstein's theory, the unique globally vacuum astrophysical solution for a spinning object
is the Kerr geometry. It depends on two parameters only: its mass $M$ and angular momentum
$J=GM^2\chi/c$, with $G$ Newton's constant, $c$ the speed of light and $|\chi|\leq1$ a pure number. 
The compact, dark objects in our universe could depart from the Kerr geometry in two distinct ways. The near-horizon structure might change significantly, while retaining the horizon~\cite{Giddings:2013kcj,Giddings:2014nla,Giddings:2017mym}.
In coalescing binaries, such effects can be probed by GW measurements of the quadrupole moment, the tidal absorption and deformability~\cite{Krishnendu:2017shb,Cardoso:2016oxy,Cardoso:2017cfl,Cardoso:2017cqb}, and especially the quasinormal oscillation modes of the remnant object~\cite{Berti:2009kk,Berti:2016lat}. 
Here we explore a second (and more subtle) scenario, where
the geometry is nearly everywhere the same as that of a BH, but the horizon is absent.
Two smoking-gun effects arise in this scenario. First, the late-time ringdown consists generically of a series of slowly damped ``echoes''~\cite{Barausse:2014tra,Cardoso:2016rao,Cardoso:2016oxy,Cardoso:2017cqb}. Furthermore, by working as a one-way membrane, horizons act as a sink for external fluctuations, including those inside the {\it ergoregion}, where negative-energy states are possible~\cite{Penrose:1969pc,Brito:2015oca}. Such states are typically associated with instabilities: their existence allows scattering waves to be amplified, i.e. positive-energy perturbations can be
produced, which can travel out of the ergoregion. Energy conservation then requires the negative-energy states inside the ergoregion to grow.
In a BH, this piling up can be avoided by dumping the negative energy into the horizon, thus stabilizing the object.
In the absence of a horizon, instead, this process leads to an exponential cascade.
As a consequence, spinning BHs are linearly stable, but any horizonless object sufficiently similar to a rotating BH is unstable~\cite{1978CMaPh..63..243F,Moschidis:2016zjy,Cardoso:2007az,Cardoso:2008kj}.

\subsection{Canonical model: perfectly reflecting surface}
%
\begin{figure}[t]
\begin{center}
\begin{tabular}{c}
  \includegraphics[width=\columnwidth]{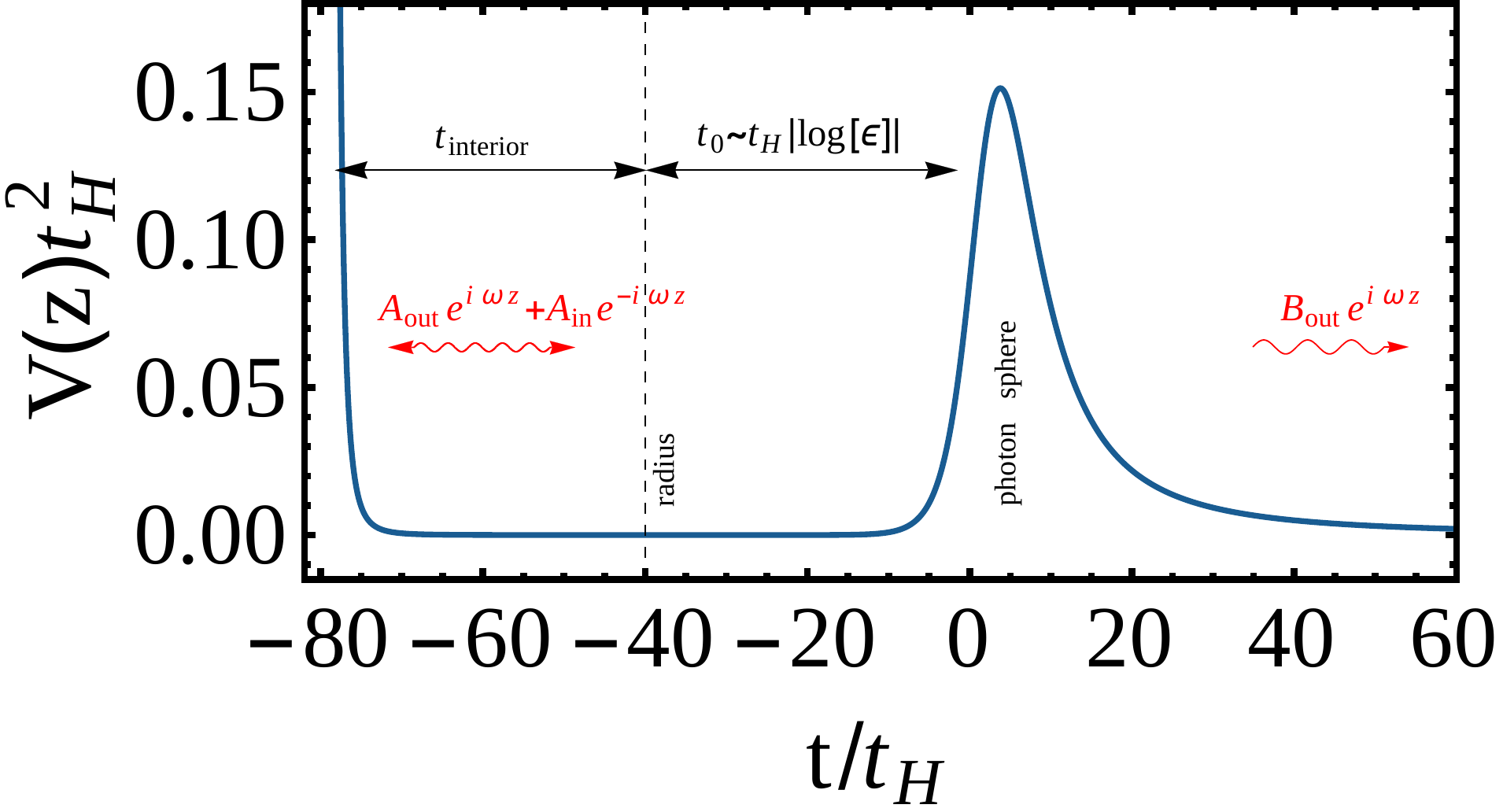}\\
\end{tabular}
\caption{Schematic potential for a non-spinning ultracompact object, as a function the tortoise radial coordinate (in practice, the coordinate time $t$ of a photon). For object radii $r_0\sim r_+$, the radiation travel time from the photosphere to the surface scales approximately as $t_0\sim t_H|\log\epsilon|$. The travel time from the surface to the interior is parametrized as $t_{\rm interior}$.  
\label{fig:potential} 
}
\end{center}
\end{figure}
We start by the simplest model of horizonless geometries: a compact body whose exterior
is described by the Kerr metric, and with a perfectly reflective surface. 
This spacetime defines a natural cavity, i.e. the region between the object's surface and the potential barrier for massless particles (the ``photon sphere'') [see Fig.~\ref{fig:potential}].
In this cavity, negative-energy modes, and thus instabilities, can be excited~\cite{Vilenkin:1978uc,Cardoso:2017njb}.
The dynamics is controlled by two parameters: the size of the cavity and the object's angular velocity, which determine 
how fast the instability grows. The cavity size can be parametrized by the light travel time $t_0$ (as observed at infinity) between the photon sphere and the object's surface~\cite{Brito:2015oca,Cardoso:2016rao,Cardoso:2017cfl,Cardoso:2017cqb}. The timescale $t_0$ also defines a set of possible modes, with fundamental frequencies
$\omega=\omega_R+i \omega_I$ and  $\omega_R\sim \pi/t_0$. The instability is controlled by the amplification factor $|{\cal A}|^2$ of the ergoregion at this frequency~\cite{Brito:2015oca}, i.e. $\omega_I\sim |{\cal A}|^2/t_0$. This follows from a very generic ``bounce-and-amplify'' argument, which was shown to accurately describe specific models~\cite{Brito:2015oca,Cardoso:2017njb}.

A scalar, electromagnetic, or gravitational perturbation in 
such a geometry grows exponentially on a timescale $\tau\equiv 
1/\omega_I$. The characteristic 
unstable modes can be computed numerically
and agree well with bounce-and-amplify estimates~\cite{Cardoso:2017njb,Maggio:2018ivz}. When 
$t_0\gg t_H\equiv {GM}/{c^3}$, these modes are well described 
by~\cite{Vilenkin:1978uc,Cardoso:2008kj,Cardoso:2017njb,Maggio:2017ivp,
Maggio:2018ivz}
\begin{eqnarray}
\omega_R &\simeq&  -\frac{q\pi}{2t_0}+m\Omega 
\,, \label{omegaR}\\
\omega_I &\simeq& -\beta_{ls}\frac{t_H}{t_0}(1+\Delta){\Delta^{2l}} 
\left(2\omega_R t_H\right)^{2l+1} (\omega_R-m\Omega)\,, \label{omegaI}
\end{eqnarray}
where $\Delta=\sqrt{1-\chi^2}$, $\Omega(\chi)$ is the object's angular 
velocity, ${\beta_{ls}^{1/2}}={(l-s)!(l+s)!}/[{(2l)!(2l+1)!!}]$ for a spin-$s$ 
perturbation with angular number $l$~\cite{Starobinskij2,Brito:2015oca}, and $q$ is a positive odd
(even) integer for Dirichlet (Neumann) boundary conditions at the 
surface~\cite{Maggio:2018ivz}.
Thus, the spacetime is unstable for $\omega_R(\omega_R-m\Omega)<0$ (i.e., in the 
superradiant regime~\cite{Brito:2015oca}), on a timescale
$\tau\equiv 1/\omega_I$. Here we consider the dominant 
gravitational mode ($l=m=2$) to estimate the background of GWs.

Equations~\eqref{omegaR} and~\eqref{omegaI} are valid for any object able to completely reflect the incoming radiation. In particular, 
if the object's surface sits at a constant (Boyer-Lindquist) radius
\begin{equation}
r_0=r_+(1+\epsilon)\,,
\end{equation}
where $r_+=GM(1+\Delta)/c^2$ is the location of the (would-be) event horizon in these coordinates, then the travel time reads~\cite{Cardoso:2016rao,Cardoso:2016oxy,Cardoso:2017cqb}
\be
t_0^{\rm canonical} \sim t_H |\log\epsilon| \,.\label{t0_canonical}
\ee
Several different arguments about the magnitude of $\epsilon$ can be made. If quantum-gravity effects become important at
Planck timescales $t_P=\sqrt{\hbar G/c^5}$, it is natural to set $\epsilon=t_P/t_H\sim 10^{-39}-10^{-46}$ for stellar-mass to supermassive dark objects. 
These objects were dubbed {\em ClePhOs} in Ref.~\cite{Cardoso:2017njb}, and are impossible to rule out in practice via electromagnetic observations~\cite{Cardoso:2017cqb,Cardoso:2017njb}.

\subsection{Modelling the interior}

The above description effectively decouples the outside geometry from the inside, and is accurate when the flux across the surface vanishes. 
However, some models may have important transmittance. There are thus three different scenarios that need to be discussed in the general case:

\noindent {\bf i.} The object does not dissipate, and the light travel time 
inside the object is small (i.e., $t_{\rm interior}\sim t_H\ll t_0$) . This 
situation describes most of the known models available in the literature, including gravastars~\cite{Mazur:2004fk,Cardoso:2016oxy}, for which the Shapiro delay dominates the travel time. In such a case, the 
geometric center of the star effectively works
as a perfectly reflecting mirror (i.e. ingoing radiation from one side exits on the other side with negligible delay), and the previous results~\eqref{omegaR}-\eqref{t0_canonical} still apply.

\noindent {\bf ii.} The object does not dissipate, and $t_{\rm interior}\gg 
t_0$. This model includes, for instance, ultracompact incompressible 
stars. These have moderate redshift and Shapiro delays in their exterior, but the light travel time 
in their interior can be very long~\cite{Cardoso:2017njb,Pani:2018flj}. For 
these objects, a model of the interior
is necessary, because even though all ingoing radiation will exit on the other 
side, delays/scattering due to the propagation in the interior will be 
important. To describe this case in a model-independent way, we assume 
that~\eqref{omegaR}-\eqref{omegaI} continue to apply, and we promote $t_0$ to a free parameter, without assuming 
Eq.~\eqref{t0_canonical}.

\noindent {\bf iii.} The object dissipates radiation in its interior. In this case, the instability may be completely quenched if the absorption rate is large~\cite{Maggio:2017ivp}. For highly spinning objects, this requires at least $0.4\%$ absorption rate for scalar fields, but up to $100\%$ absorption rate for gravitational perturbations and almost maximal spins~\cite{Press:1972zz,Brito:2015oca,Maggio:2018ivz}. While these numbers reduce to $\lesssim 0.1\%$ for spins $\chi \lesssim 0.7$, they are still several orders of magnitude larger than achievable with viscosity from nuclear matter~\cite{Maggio:2017ivp}. 

Based on the above arguments, we expect the following results to cover all relevant models of BH mimickers.

\subsection{Evolution of the instability}
The evolution of the object's mass and angular momentum, under energy and angular momentum losses, can be computed within the adiabatic approximation (because $\tau\gg t_H$)~\cite{Brito:2014wla}.
The unstable mode is simply draining energy and angular momentum from the object, which we assume to have an equation
of state such that $\epsilon={\rm const}$ during the evolution.
From energy and angular momentum conservation, the evolution equations for each mode read
\be\label{evol_equations}
\dot E = \dot{E}_0\,e^{2\omega_I t}\,, \quad\dot{M}=-\dot E/c^2\,,\quad \dot{J}=-m{\dot E}/{\omega_R}\,,
\ee
where $\dot{E}_0$ encodes the initial preturbation of the (unstable) system. Since the instability is exponential, the overall evolution is insensitive to the precise value of these initial conditions. The equations above are valid for a monochromatic mode in a generic stationary and axisymmetric background. The energy flux can be written as $ dE/df = \dot E /\dot f$,
%
%
where $f=\omega_R/(2\pi)$ is the frequency associated with the mode. From the evolution equations, we can evaluate $\dot M$ and $\dot J$ and, in turn, $\dot \omega_R$. To leading order in the $\epsilon\to0$ limit, we obtain $\omega_R\sim m\Omega$, 
and
\begin{equation}\label{dEdf}
 \frac{dE}{df} \sim \frac{16G^2\pi^2}{m^2c^4} f M^3\,,  
\end{equation}
valid for any angular numbers $l$ and $m$.  In the same limit, the critical value of spin above which the ergoregion instability occurs reads~\cite{Maggio:2017ivp,Maggio:2018ivz}
%
%
\begin{equation}
\chi_{\rm crit} \approx\frac{\pi q}{m |\log(\epsilon)|} \approx \frac{0.035\,q}{m} \log^{-1}\left(\frac{\epsilon}{10^{-40}}\right)\,.
\end{equation}

Thus, if $\chi(t=0)>\chi_{\rm crit}$, the instability removes energy and angular momentum until superradiance is saturated, i.e. $\chi(t\gg\tau)\to\chi_{\rm crit}$.
Note that the small spin value $\chi_{\rm crit}$ is compatible with the low 
measured spins of inspiralling compact objects detected via GWs so 
far~\cite{TheLIGOScientific:2016pea}. Since we are interested in gravitational 
perturbations, when solving the evolution equations~\eqref{evol_equations} and 
computing the energy flux~\eqref{dEdf} we only consider the dominant $l=m=2$ 
mode and neglect higher-modes. For each initial spin $\chi$ and travel 
time $t_0$ we only consider the dominant $q$-mode associated with the shortest 
instability timescale. Note also that the above analysis assumes that the 
backreaction of these fields on the geometry is negligible. Our results indicate 
that those are always reliable approximations.
\subsection{GW stochastic background}
A population of GW sources too far and/or weak to be detected individually may still
give rise to a ``stochastic'' background 
detectable by a network of interferometers, e.g. the LIGO/Virgo network,
sensitive to frequency ranging from $\sim 10$ to $\sim 100$ Hz~\cite{TheLIGOScientific:2014jea,aVIRGO}; 
the Pulsar-Timing-Array experiments~\cite{2013CQGra..30v4009K,2013CQGra..30v4007H,2013CQGra..30v4008M,2010CQGra..27h4013H,2013CQGra..30v4010M}, 
which are already constraining backgrounds at frequencies $\sim10^{-9}-10^{-6}$ Hz; and
the future LISA constellation~\cite{Audley:2017drz}, 
which will be sensitive to frequencies between $10^{-6}$ Hz and $\sim 1$ Hz.
The background is produced by the incoherent
superposition, at the detector, of the GW signals from all the unresolved sources in the population.
The background can be characterized either by \textit{(i)} its (dimensionless) 
energy spectrum
\begin{equation}
\Omega_{\rm gw}(f_o)=\frac{1}{\rho_c}\frac{d\rho_{\rm gw}}{d\ln f_o}\,, \label{OmegaGW}
\end{equation}
($\rho_{\rm gw}$ being the background's energy density, $f_o$ the frequency measured at the detector
and $\rho_c$ the critical density of the Universe at the present time),
obtained by summing the energies emitted by all the unresolved sources in a 
given frequency bin~\cite{Brito:2017zvb}; 
or \textit{(ii)} directly by the characteristic strain $h_c(f_o)$ observed in the detector,
which can be obtained summing in quadrature (and binning in frequency) the strain amplitudes of all the unresolved sources~\cite{Sesana:2008mz}.
The two quantities are related by 
$\Omega_{\rm gw}(f_o)={2\pi^2} \left[f_o h_c(f_o)\right]^2/{(3H_0^2)}$, 
where $H_0\approx 68 {\rm km/(s\, Mpc)}$
is the Hubble rate. 
While these two ways of computing the background signal are equivalent~\cite{Sesana:2008mz}, we have
implemented both as a consistency check of our results. 
This also allows us to check that
the number of sources contributing in each frequency bin is typically large as long as the bin size is $\gtrsim 0.01$ dex in the LISA band (which ensures
that the number of sources contributing 99\% of the signal in each bin ranges from 
thousands to millions). In the LIGO band, sources are even more numerous: frequency bins $\gtrsim 0.01$ dex
yield $10^9$--$10^{14}$ sources contributing 99\% of the signal in each bin. (These are mostly extragalactic sources, as Galactic ones give a negligible contribution to the background.)
This in turn implies, in particular, that the background is
expected to be smooth with that frequency binning~\cite{Sesana:2008mz}.
When computing the stochastic background of unstable exotic compact objects, the energy flux given by Eq.~\eqref{dEdf} is defined in the frequency range $f\in [f_{\rm min},f_{\rm max}]$; $f_{\rm max}$ can be computed using Eq.~\eqref{omegaR} for a given initial mass and spin of the compact object, and $f_{\rm min}$ is computed by solving the evolution equations~\eqref{evol_equations} from the formation redshift of the compact object to the present time.

\begin{figure*}[t]
\includegraphics[width=0.48\textwidth]{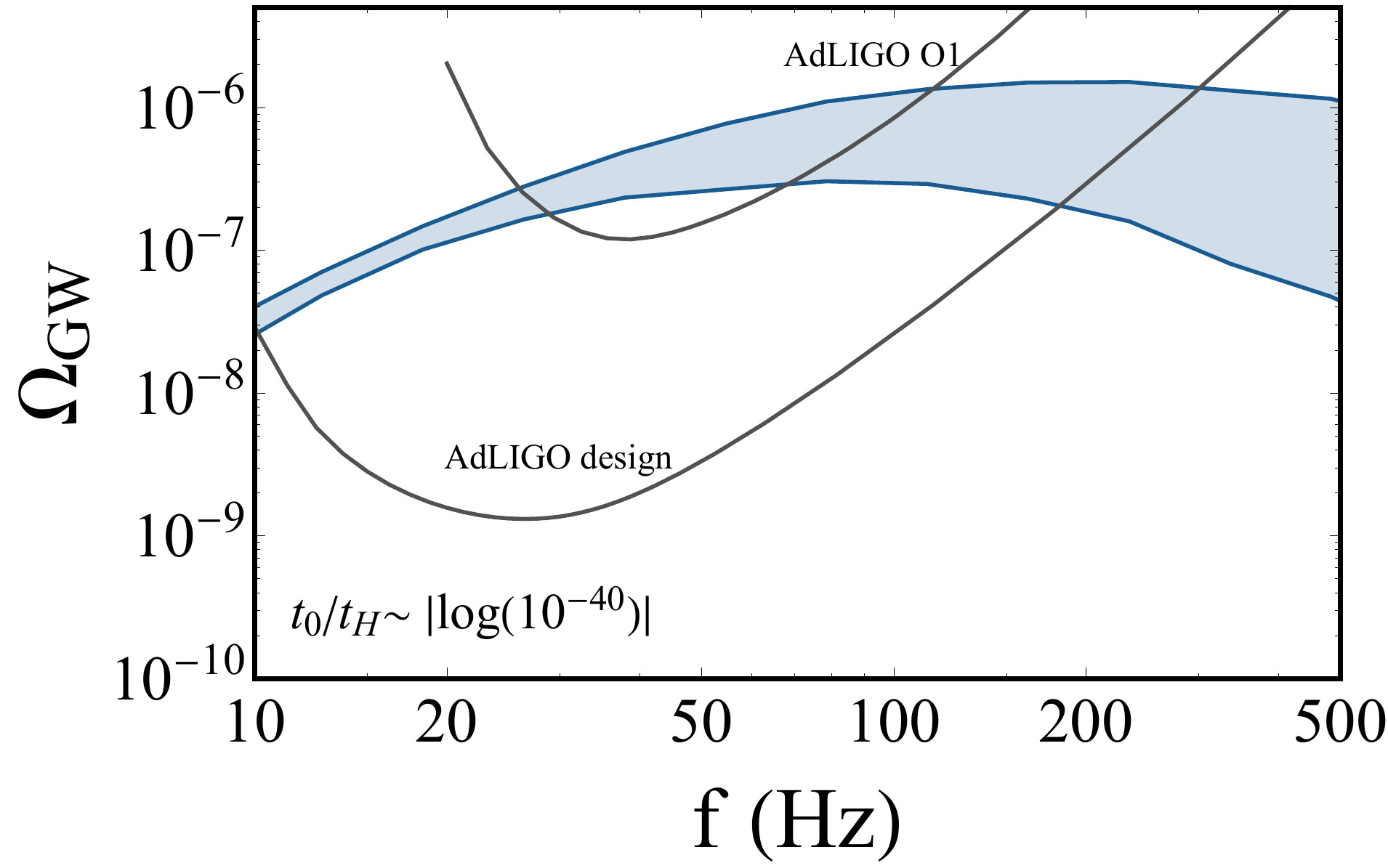}
\includegraphics[width=0.46\textwidth]{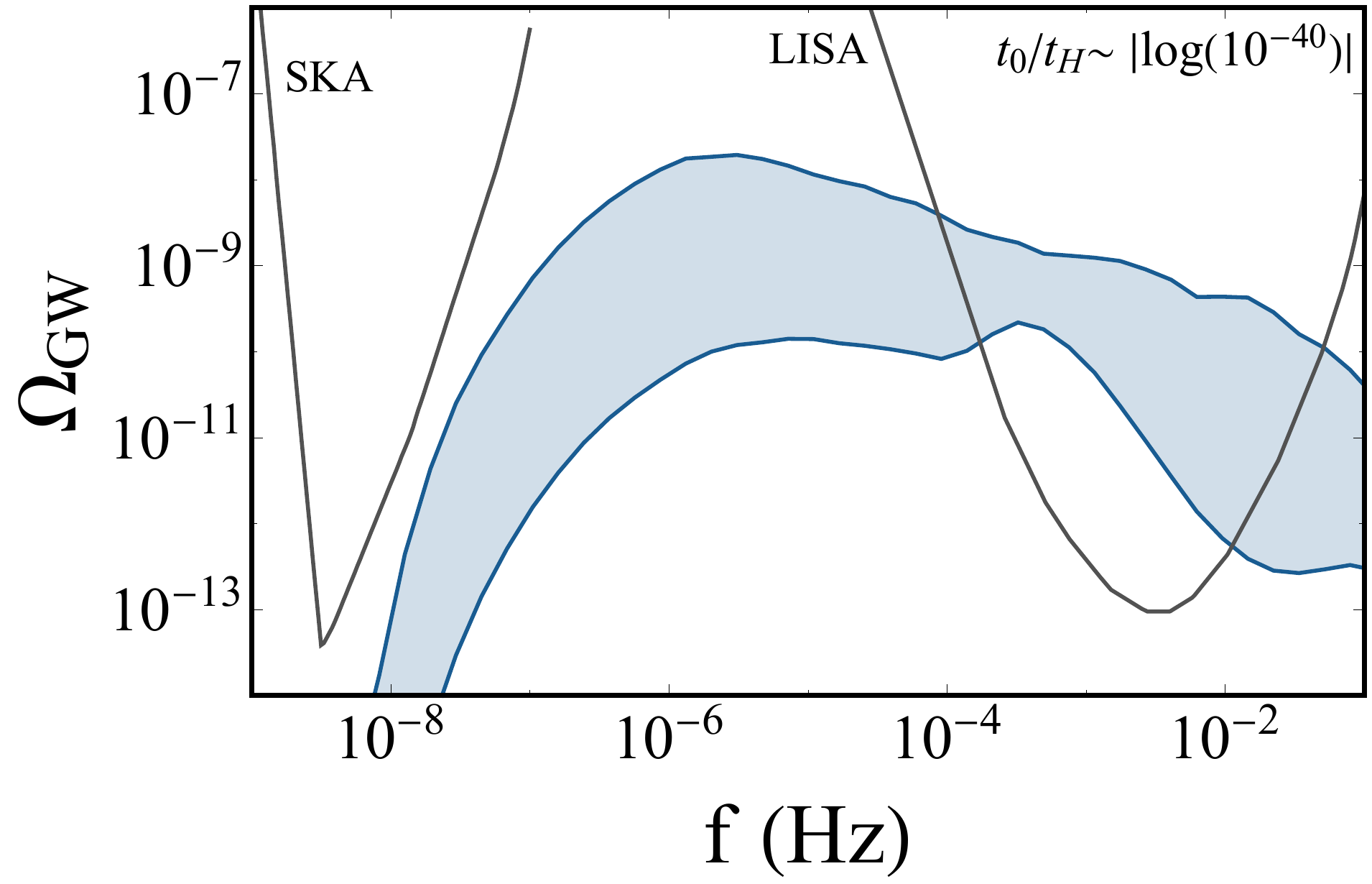}
\caption{Extragalactic stochastic background for the canonical model in the LIGO/Virgo (left panel), LISA and PTA bands (right panel). 
The blue band brackets our population models (from the most pessimistic to the most optimistic, as explained in the main text). The background depends very weakly on $\epsilon$ as long as $t_0\sim t_H| \log\epsilon | \ll 10^{10} t_H$, so here we show only the case $t_0\sim t_H | \log 10^{-40}|$. The black lines are the power-law integrated curves of~\cite{Thrane:2013oya}, computed using noise PSDs for LISA with one year of observation time~\cite{Audley:2017drz}, LIGO's first observing runs (O1), LIGO at design sensitivity as described in~\cite{TheLIGOScientific:2016dpb}, and an SKA-based pulsar timing array as described in \cite{Dvorkin:2017vvm}. By definition, $\rho_{\rm stoch}> 2$ ($\rho_{\rm stoch}= 2$) when a power-law spectrum intersects (is tangent to) a power-law integrated curve.
\label{fig:background}
}
\end{figure*}
\begin{figure*}[t]
\includegraphics[width=0.48\textwidth]{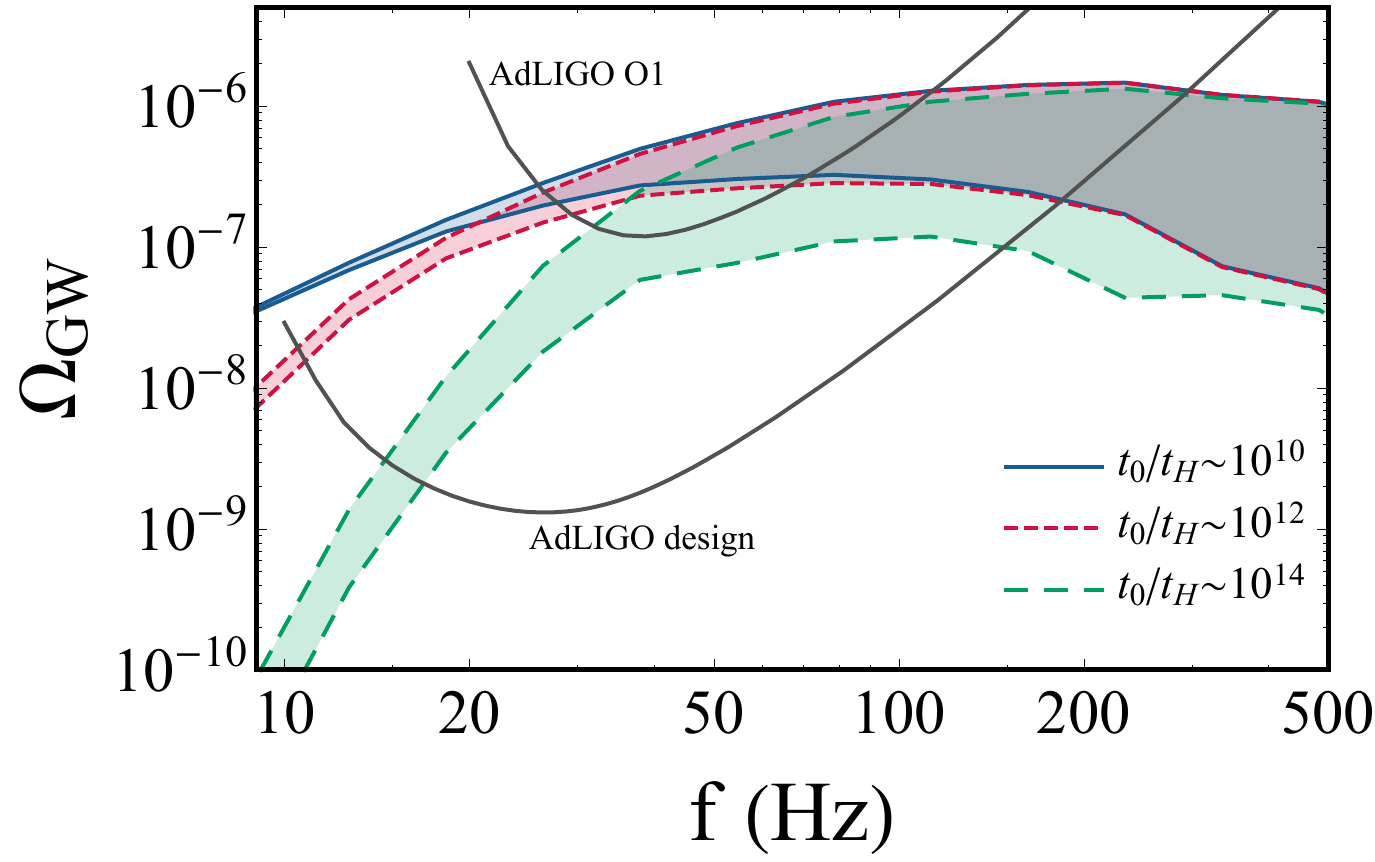}
\includegraphics[width=0.48\textwidth]{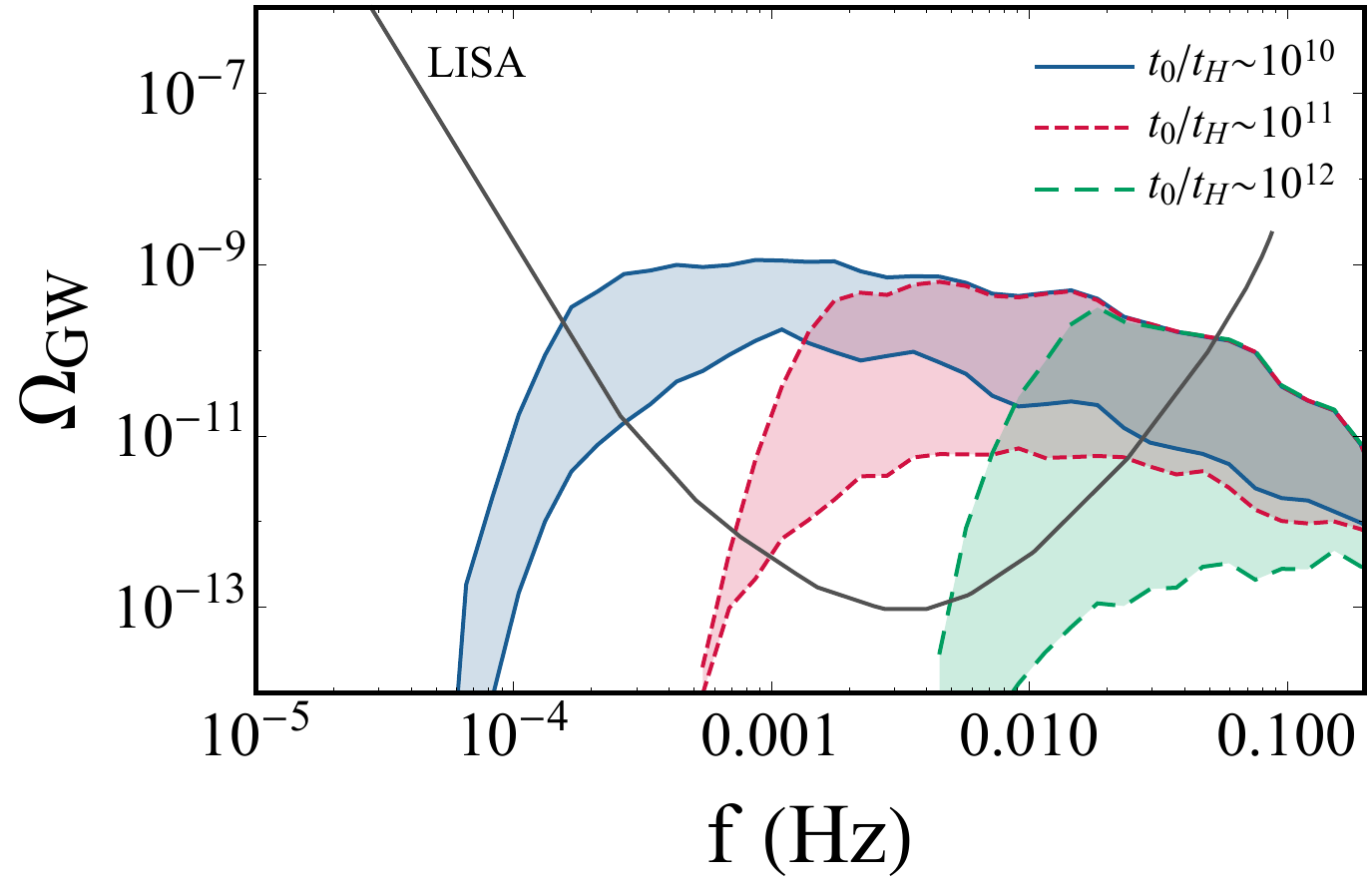}  
\caption{
Same as in Fig.~\ref{fig:background}, but for an agnostic model for the compact-object (dissipationless) interior, where the light travel time $t_0$ between the light ring and the surface is a free parameter.
 \label{fig:LIGOextrabackground}
}
\end{figure*}
For the astrophysical populations of isolated BHs, we adopt the same models as in~\cite{Brito:2017zvb,Brito:2017wnc}. We
assume that all such astrophysical BH candidates are actually BH mimickers.
For stellar-origin BHs we account for both Galactic and 
extragalactic BHs that form from the core collapse of massive ($\gtrsim 20 M_\odot$) stars, by tracking the 
cosmic star formation history and the metallicity evolution of the Universe~\cite{2016MNRAS.461.3877D}. We assume a uniform distribution for the initial spins with $\chi\in[0,1]$ as the most optimistic and $\chi\in[0,0.5]$ as the most pessimistic scenario. 
For the massive ($\sim 10^4-10^7 M_\odot$) and supermassive ($\sim 10^8-10^{10} M_\odot$) BHs that emit respectively in the LISA and PTA bands, we
adopt the semi-analytic galaxy-formation model of Ref.~\cite{Barausse:2012fy} (with later incremental improvements
described in~\cite{Sesana:2014bea,newpaper,letter}), which follows the formation of these objects from
their high-redshift seeds and their growth by accretion and mergers. This growth is 
triggered in turn by the synergic co-evolution of the BHs with their host galaxies, of which we evolve both
the various baryonic components and the dark-matter halos. This model is optimistic since it predicts a spin distribution skewed towards large spins, at least at low masses. To include astrophysical uncertainties in our computation, we also consider models in between our most optimistic and most pessimistic assumptions as described in Ref.~\cite{Brito:2017zvb} (see Section~III therein).

\section{Results}
Our main results for the GW stochastic background from exotic compact objects are shown in Fig.~\ref{fig:background} in the frequency bands relevant for LIGO/Virgo (left panel) and for LISA/an SKA-based pulsar timing arrays (right panel). The left panel suggests that the absence of a stochastic background in LIGO O1 already rules out our canonical model
even for conservative spin distributions, while LIGO at design sensitivity will be able to rule out our canonical model even in more pessimistic scenarios than those assumed here, e.g. even if all BH-like objects had initial spin $\chi<0.2$.  Similar results apply in the LISA band, whereas the stochastic signal is too small to be detectable by pulsar timing arrays, even in the SKA era.\footnote{This is because the frequency given by Eq.~\eqref{omegaR} is in the range of pulsar timing arrays
only for BH mimickers of masses $\gtrsim 10^{11} M_{\odot}$, where astrophysical BH candidates are expected to be extremely rare (if any exist)~\cite{2018MNRAS.474.1342M}, 
or for systems of smaller masses but also lower spins, which either emit GWs very weakly or are stable. These latter sources do indeed
produce the small background visible in Fig.~\ref{fig:background}. Moreover, even if BH mimickers with masses $\gtrsim 10^{11} M_{\odot}$ existed,
Eq.~\eqref{omegaI} gives instability timescales larger than the Hubble time, i.e. these systems are effectively stable.}

The level of the stochastic background shown in Fig.~\ref{fig:background} can also be understood with an approximate analytic 
calculation~\cite{Brito:2017wnc}. The BH-mimicker mass fraction lost to GWs due to
superradiance is $F_{\rm sr}\sim {\cal O}(1\%)$~\cite{Brito:2017zvb,2017arXiv171200784F}.
Because the signal spans about a decade in frequency [c.f. Eq.~\eqref{omegaR}],  $\Delta \ln f \sim 1$,
and $\Omega_{\rm GW,\,sr} =({1}/{\rho_{\rm c}})({d\rho_{\rm GW}}/{d\ln f})
\sim F_{\rm sr}{\rho_{\rm BH}}/{\rho_{\rm c}}$, with $\rho_{\rm GW}$
and $\rho_{\rm BH}$  the GW and BH-mimicker energy densities.
In the mass
range $10^4-10^7 M_\odot$ relevant for LISA,
$\rho_{\rm BH}\sim {\cal O}( 10^4) M_{\odot}/{\rm Mpc}^3$, which
gives $\Omega^{\rm LISA}_{\rm GW,\,sr} \sim 10^{-9}$.
To estimate the background in the LIGO band, note that the background from ordinary
BH binaries is $\Omega_{\rm GW,\,bin}\sim \eta_{\rm GW} F_{\rm m} \rho_{\rm
  BH}/\rho_c$, with  $\eta_{\rm GW}\sim {\cal O}(1\%)$ the GW emission efficiency for BH binaries~\cite{Barausse:2012qz}, and
$F_{\rm m} \sim {\cal O}(1\%)$~\cite{2016MNRAS.461.3877D} the
fraction of stellar-mass BHs in merging binaries. This gives
$\Omega_{\rm GW,\,sr}/\Omega_{\rm GW,\,bin}\sim F_{\rm sr}/(\eta_{\rm GW}
F_{\rm m})\sim 10^2$, and  because the LIGO O1 results imply  
$\Omega_{\rm GW,\,bin}\lesssim 10^{-9} - 10^{-8}$~\cite{TheLIGOScientific:2016wyq,TheLIGOScientific:2016dpb}, we obtain $\Omega^{\rm LIGO}_{\rm GW,\,sr}\lesssim10^{-7}-10^{-6}$.

As shown in Fig.~\ref{fig:LIGOextrabackground}, LIGO/Virgo and LISA are also able to place model-independent constraints on the stochastic signal from exotic compact objects. At design sensitivity, LIGO/Virgo can detect or rule out any model with $t_0<10^{14} t_H$, whereas LISA can go as far as $t_0<10^{12}t_H$. In other words, objects across which light takes $10^{14}$ or less dynamical timescales to travel are ruled out. Finally, we note that although LIGO/Virgo rule out a wider range of the parameter space compared to LISA, it is still interesting to consider the constraints from both detectors since they probe different BH populations.

\section{Discussion}

Our results suggest that the current upper limits on the stochastic background from LIGO O1 already rule out the simplest models of BH mimickers at the Planck scale,
setting the strongest constraints to date on exotic alternatives to BHs. The most relevant parameter for our analysis is the light travel time within the object, $t_0$. LIGO/Virgo (LISA) can potentially rule out models where $t_0<10^{14}t_H$ ($t_0<10^{12}t_H$). These results are not significantly dependent on the astrophysical uncertainties of the extragalactic BH distributions, and even if exotic compact objects are produced all at low spin, the order of magnitude of our constraints is unaffected.

We conclude that all present and future models of exotic compact objects should either conform with $t_0\gg 10^{14} t_H$ {\it or} represent at most a fraction $X$ of the compact-object
population, the remaining being BHs. Since Eq.~\eqref{OmegaGW} scales linearly with $X$, Fig.~\ref{fig:background}
implies that the O1 upper limits impose $X<50\%$ even if all BH-like objects are formed with low spins. At design sensitivity, these constraints could improve to $X<1\%$.

It might be possible to evade these constraints, by incorporating some (exotic and still unclear) mechanism quenching the ergoregion instability, e.g. absorption rates several orders of magnitude larger than those of neutron stars~\cite{Maggio:2017ivp}. 
While such quenching mechanism might result in thermal or quasi-thermal electromagnetic radiation (which can be constrained  by electromagnetic observations of BH candidates~\cite{Broderick:2005xa,Broderick:2007ek}), quantum-dressed BH mimickers might evade such constraints by trapping thermal energy in their interiors for very long timescales~\cite{Cardoso:2017cqb,Cardoso:2017njb}.

Our results also imply that in the simplest models of non-dissipative exotic ultracompact objects, GW echoes~\cite{Cardoso:2016rao,Cardoso:2016oxy,Cardoso:2017cqb} can appear in the post-merger phase only after a delay time $\tau_{\rm echo}\sim t_0\gg 10^{14} t_H\approx  10^{10}[M/(20M_\odot)]{\rm s}$, which is orders of magnitude longer than what was claimed to be present in LIGO/Virgo data~\cite{Abedi:2017isz,Abedi:2018npz,Abedi:2018pst} (see also~\cite{Westerweck:2017hus}). The latter claims would not be in tension with our bounds on the GW stochastic background only if one postulates exotic objects that are dissipative enough to absorb at least ${\cal O}(0.1)\%$ of gravitational radiation (which is several orders of magnitude more than what typically achievable with nuclear matter).
In this case, the stochastic background from a population of ``echoing'' merger remnants might still be detectable~\cite{Du:2018cmp}. 

\begin{figure}[t]
\includegraphics[width=0.48\textwidth]{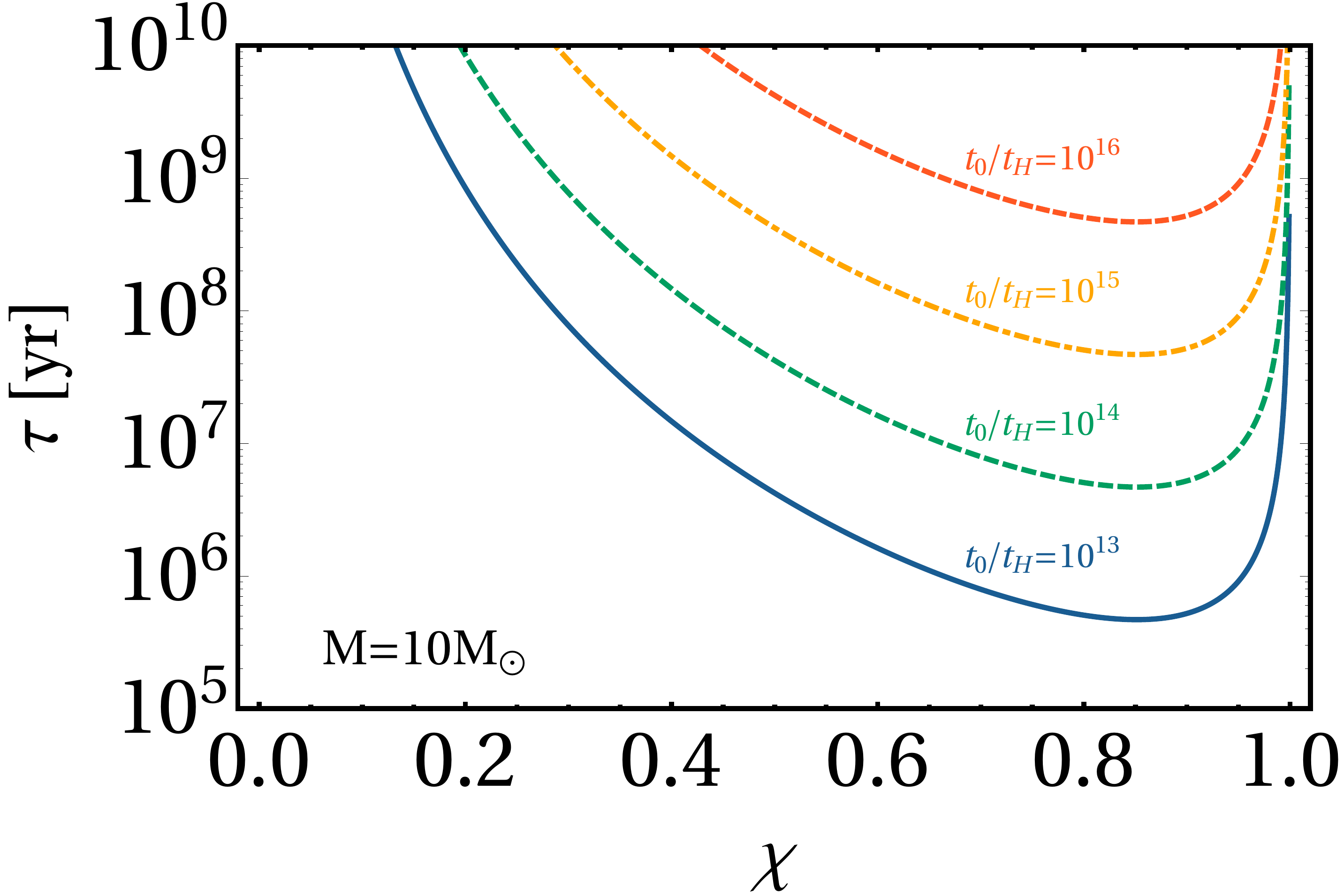}
\caption{Instability timescale for $l=m=2$ gravitational perturbations as 
function of 
the BH spin $\chi$, for different values of $t_0$. 
 \label{newfig}
}
\end{figure}

Finally, our constraints are stronger than those one might infer from BH 
spin measurements in X-ray binaries (see e.g.~\cite{Middleton:2015osa}). 
Indeed, those observations can only rule out instability timescales shorter than 
the BH age (in the more likely case in which the spin is 
natal~\cite{McClintock:2011zq}, e.g. from supernova explosions)  or the Salpeter 
timescale $t_{\rm S}\sim 10^7$ yr (if the BH spin is produced by accretion), 
otherwise the BH would have no time to spin down under the effect of the 
instability. However, the BH age is hard to reconstruct from the observed source properties,
and in the few high-spin X-ray binaries where it was obtained via population synthesis modelling,
it is of a few Myr~\cite{2012ApJ...747..111W,2010Natur.468...77V}. Therefore, obtaining exact bounds on the 
instability is tricky in the case of natal BH spins.
Even if  BH spins in X-ray binaries were
accretion-produced, which is unlikely~\cite{McClintock:2011zq,2012ApJ...747..111W}, 
the obtained bound would simply be $t_{\rm inst}\gtrsim t_{\rm S}$, 
which would still be marginally weaker than the constraints presented in this paper (c.f. Fig. 
\ref{newfig}).

Let us also add that spin measurements in X-ray binaries are likely affected by
unknown systematics (in several cases different techniques yield different 
results, c.f. Table 1 in \cite{Middleton:2015osa}). More importantly, the very 
existence of the ergoregion instability in BHs surrounded by gas has never been 
investigated in detail, and the backreaction of the disk mass and angular 
momentum on the geometry, as well as the viscosity of the gas, may change the 
character and timescale of the instability.

Finally, let us comment on the relation between this work and~\cite{2017arXiv171200784F}, which computes the stochastic background from a hypothetical spin-down mechanism of BH merger remnants (in the LIGO/Virgo band only).
Besides performing more realistic calculation of the spectrum $dE/df$ due to the ergoregion instability,
our work crucially differs from \cite{2017arXiv171200784F} as we compute the
background from {\em all} BH mimickers in the Universe (including isolated ones), unlike \cite{2017arXiv171200784F} which only accounted for objects resulting from binary mergers.
Since isolated compact objects are expected to be $1/F_{\rm m}\sim 100$ times more numerous than merging binaries~\cite{2016MNRAS.461.3877D}, our background level is about 100 times larger, which allows constraining BH alternatives with current LIGO/Virgo data.

\begin{acknowledgments}
We thank Andrew Matas for a careful reading of the manuscript.  We thank Jing 
Ren and Elisa Maggio for useful discussions.
V.C. acknowledges financial support provided under the European Union's H2020 ERC Consolidator Grant ``Matter and strong-field gravity: New frontiers in Einstein's theory'' grant agreement no. MaGRaTh--646597.
PP acknowledges financial support provided under the European Union's H2020 ERC, Starting Grant agreement
no.~DarkGRA--757480.
This project has received funding from the European Union's Horizon 2020 research and innovation programme under the
Marie Sklodowska-Curie grant agreement No 690904.
The authors would like to acknowledge networking support by the COST Action CA16104.
\end{acknowledgments}

\bibliographystyle{apsrev4}
\bibliography{Ref}

\end{document}